\documentclass[10pt,twocolumn,letterpaper]{article}

\usepackage{cvpr}
\usepackage{times}
\usepackage{graphicx}
\usepackage{amsmath}
\usepackage{amssymb}
\usepackage{subfigure}
\usepackage{epsfig}
\usepackage{verbatim}
\usepackage{multirow}
\usepackage{color}
\usepackage[font={small}]{caption}
\usepackage{booktabs}
\usepackage{xspace}
\usepackage{makecell}
\usepackage{wrapfig}
\usepackage{floatrow,makecell}
\usepackage{amsbsy}

\newenvironment{packed_enumerate}{
\begin{enumerate}
  \setlength{\itemsep}{0pt}
  \setlength{\parskip}{0pt}
  \setlength{\parsep}{0pt}
\vspace{-4pt}
}{\end{enumerate}}

\usepackage[pagebackref=true,breaklinks=true,letterpaper=true,colorlinks,bookmarks=false]{hyperref}

\cvprfinalcopy 

\def\cvprPaperID{187} 

\ifcvprfinal\fi
\begin{document}

\title{Understanding Image Virality}

\author{Arturo Deza\\
UC Santa Barbara\\
{\tt\small deza@dyns.ucsb.edu}
\and
Devi Parikh\\
Virginia Tech\\
{\tt\small parikh@vt.edu}
}

\maketitle

\begin{abstract}
Virality of online content on social networking websites is an important but esoteric phenomenon often studied in fields like marketing, 
psychology and data mining. In this paper we study viral images from a computer vision perspective. We introduce three new
image datasets from \emph{Reddit}\footnote{\url{www.reddit.com}, Reddit is considered the main engine of virality around the world, and is 
ranked $24^{th}$ among the top 
sites on the web by Alexa (\url{www.alexa.com}) as of March 2015} 
and define a virality score using \emph{Reddit} metadata. 
We train classifiers with state-of-the-art image features to predict virality of individual images, relative virality in pairs of images, 
and the dominant topic of a viral image. We also compare
machine performance to human performance on these tasks. 
We find that computers perform poorly with low level features, and high level information is critical for 
predicting virality. We encode semantic information through relative attributes.
We identify the 5 key visual attributes that correlate with virality. 
We create an attribute-based characterization of images that can predict 
relative virality with $68.10\%$ accuracy (SVM+Deep Relative Attributes) --better than humans at $60.12\%$.
Finally, we study how human prediction of image virality varies with different ``contexts'' in which the images are viewed,
such as the influence of neighbouring images, images recently viewed, as well as the image title or 
caption. This work is a first step in understanding 
the complex but important phenomenon of image virality.
Our datasets and annotations will be made publicly available.
\end{abstract}

\vspace{-10pt}
\section{Introduction}


What graphic should I use to make a new startup more eye-catching than Instagram? Which image caption will 
help spread an under-represented shocking news? Should I put an image of a cat in my YouTube video if I want 
millions of views? These questions plague professionals and regular internet users on a daily basis. Impact of 
advertisements, marketing strategies, political campaigns, non-profit organizations, social causes, authors 
and photographers, to name a few, hinges on their ability to reach and be noticed by a large number of people. 
Understanding what makes content viral has thus been studied extensively by marketing researchers~\cite
{berger2011drives,berger2011arousal,chen2012and,berger2013contagious}.

\begin{figure}[t]
\begin{center}
\subfigure[Example viral images.]{
	\includegraphics[scale=0.18]{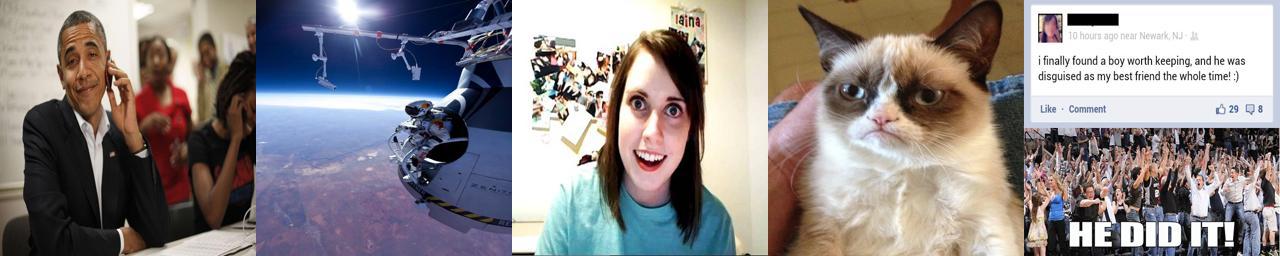}
    \label{fig:viral_exem}
}
\subfigure[Example non-viral images.]{
	\includegraphics[scale=0.18]{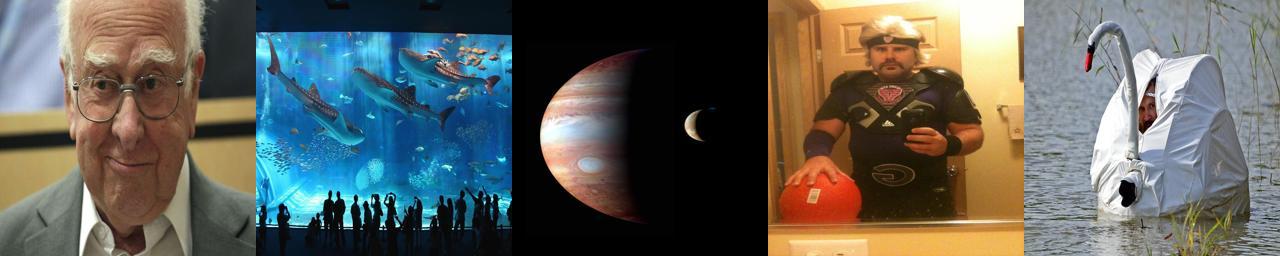}
    \label{fig:non_viral_exem}
}
\end{center}
\caption{
Top: Images with high viral scores in our dataset depict internet ``celebrity'' memes ex. 
``Grumpy Cat''; 
Bottom: Images with low viral scores in our dataset. The picture of Peter Higgs (Higgs Boson) was popular, but 
was not reposted multiple times and is hence not considered viral. 
\vspace{-20pt}
}
\label{fig:Viral_nonviral_exem}
\vspace{-20pt}
\end{figure}

Many factors such as the time of day and day of week when the image was uploaded, the title used with the image, etc. 
affect whether an image goes viral or not~\cite{Leskovec}. To what extent is 
virality dependent on these external factors, and how much of the virality depends on the image content itself? 
How well 
can state-of-the-art computer vision image features and humans predict virality?
Which visual attributes correlate with image virality? 

In this paper, we address these questions. We introduce three image databases collected from Reddit and a virality score. 
Our work identifies several interesting directions for deeper investigation where computer vision techniques can be 
brought to bear on this complex problem of understanding and predicting image virality.

\vspace{-5pt}
\section{Related Work}
\vspace{-5pt}

Most existing works ~\cite{leskovec2007dynamics,barabasi2005origin,1309.2963} study how people share content on social networking sites \emph{after} it has been posted. 
They use the network dynamics soon after the content has been posted to detect an oncoming snowballing effect and predict whether the content will go viral or 
not. 
We argue that predicting virality after the content has already been posted is too late in some applications. It is not feasible for graphics designers to ``try out'' 
various designs to see if they become viral or not. In this paper, we are interested in understanding the relations 
between the content itself (even before it is posted online) and its potential to 
be viral\footnote{In fact, if the machine understands what makes an image viral, one could use ``machine teaching''~\cite{JohnsCVPR2015} to 
train humans (e.g., novice graphic designers) what viral images look like.}.

There exist several qualitative theories of the kinds of content that are likely to go viral~\cite{berger2011arousal,berger2013contagious}. Only a few works have quantitatively analyzed content, 
for instance 
Tweets~\cite{suh2010want} and New York Times articles~\cite{berger2012makes}
to predict their virality. However, in spite of them being a large part of our online experience, 
the connections between content in visual media and their virality 
has not been analyzed. This forms the focus of our work.

Virality of text data such as Tweets has been studied in \cite{nagarajan2010qualitative,suh2010want}.
The diffusion properties were found to be dependent on their content and features like embedded URL's and hashtags.
Generally, diffusion of content over networks has been studied more than the causes~\cite{1309.2963}. 
The work of Leskovec~\etal \cite{leskovec2007dynamics} models propagation of recommendations over a network of individuals through a stochastic model,
while Beutel~\etal \cite{beutel2012interacting} approach viral diffusion as an epidemiological problem.

Qualitative theories about what makes people share content have been proposed in marketing research. 
Berger~\etal~\cite{berger2011arousal,berger2012makes,berger2013contagious} for instance
postulate a set of STEPPS that suggests that social currency, triggers, ease of emotion, public 
(publicity), practical value, and stories make people share.

Analyzing viral images has received very little attention.
Guerini~\etal~\cite{1309.3908} have provided correlations between low-level visual data and popularity on a non-anonymous social network (Google+), 
as well as the links between emotion and virality~\cite{guerini2015deep} . 
Khosla~\etal~\cite{www14_khosla} recently studied image popularity measured as the number of views a photograph 
has on Flickr. 
However, both previous works~\cite{1309.3908,www14_khosla} have only extracted 
image statistics for natural photographs
(Google+, Flickr). Images and the social interactions on Reddit 
are qualitatively different (\eg many Reddit images are edited). 
In this sense, the quality of images that is most similar to ours is the concurrently introduced 
viral \emph{meme} generator of Wang~\etal, 
that combines NLP and Computer Vision (low level features)~\cite{WangWen:2015}.
However, our work delves deep into the role of intrinsic visual 
content (such as high-level image attributes), 
visual context surrounding an image, temporal contex and textual
context in image virality. 
Lakkaraju~\etal~\cite{Leskovec} analyzed the effects of time of day, day of the week, number of resubmissions, captions, category, etc. 
on the virality of an image on Reddit. However, they do not analyze the content of the image itself.

Several works in computer vision have studied complex meta-phenomenon (as opposed to understanding the ``literal'' content in 
the image such as objects, scenes, 3D layout, etc.). Isola~\etal~\cite{isola2011makes} found that some images are consistently more 
memorable than others across subjects and analyzed the image content that makes 
images memorable~\cite{isola2011understanding}. 
Image aesthetics was studied in~\cite{dhar2011high}, image emotion in~\cite{borth2013large}, and object recognition in art in~\cite{Crowley14a}. 
Importance of objects~\cite{spain2011}, attributes~\cite{turakhia2013attribute} as 
well as scenes~\cite{berg2012} as defined by the likelihood that people mention them first in descriptions of the images has also been studied. 
We study a distinct complex phenomenon of image virality.

\vspace{-5pt}
\section{Datasets and Ground Truth Virality}\label{Creating}


\subsection{Virality Score}
\vspace{-5pt}

\begin{figure}[t]
\centering
	\includegraphics[scale=0.225]{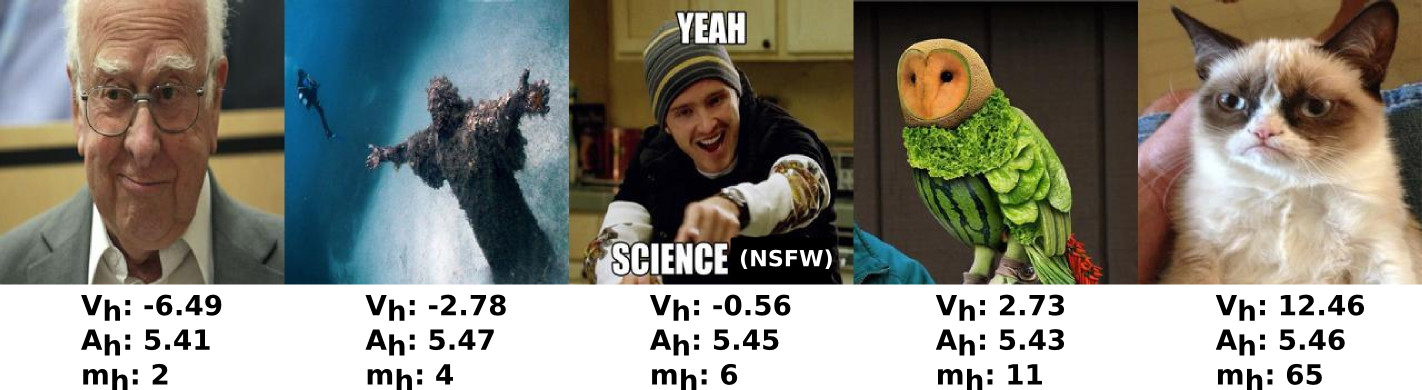}
\caption[]{Virality ($V_h$) vs. popularity ($A_h$) in images. All images have a similar popularity score, but their virality scores vary quite 
a bit. ``Grumpy Cat'' is more viral than Peter Higgs
due to number of resubmissions ($m_h$), that plays a critical role in our virality metric $V_h$. Clearly virality and popularity are two different concepts.}
\label{fig:viral_pop_sample}
\end{figure}

Reddit is the main engine of viral content around the world. Last month, it had over 170M unique visitors representing every single country. It has 
over 353K categories (subreddits) on an enormous variety of 
topics. We focus only on the image content.
These images are sometimes rare photographs, or photos depicting comical or absurd situations, or Redditors 
sharing a personal emotional moment through the photo, or expressing their political or social views through the image, and so on. Each 
image can be upvoted or downvoted by a user. Viral content tends to be resubmitted multiple times as it spreads across 
the network of users\footnote{These statistics are available through Reddit's API.}.
Viral images are thus the ones that have many upvotes, few downvotes, \emph{and} have 
been resubmitted often by different users. The latter is what differentiates virality from popularity.
Previously, Guerini~\etal defined multiple virality metrics as upvotes, shares or comments, Khosla~\etal define popularity as number of views and
Lakkaraju~\etal define popularity as number of upvotes. 
We found that the the correlation between popularity as defined by the number 
of upvotes and virality that also accounts for resubmissions (detailed definition next) is -0.02. This quantitatively 
demonstrates the distinction between these two phenomenon. See Fig.~\ref{fig:viral_pop_sample} for qualitative examples. The focus of this paper is 
to study image virality (as opposed to popularity). 

Let score $S_h^n$ be the difference between the number 
of upvotes and downvotes an image $h$ received at its $n^{th}$ resubmission to a category. 
Let $t$ be the time of the resubmission of the image and $c$ be the category (\textit{subreddit}) to which it was submitted. $\bar{S}_c^t$ is 
the average score of all submissions to category $c$ at time $t$. We define $A_h^n$ to be the ratio of the score of the image $h$ at 
resubmission $n$ to the average score of all images posted to the category in that hour~\cite{Leskovec}.

\vspace{-7pt}
\begin{equation}\label{Jure_eq}
A_h^n=\frac{S_h^n}{\bar{S}_{c}^t}
\end{equation}

We add an offset to $S_h^n$ so that the smallest score $\min_h \min_n S_h^n$ is 0.  
We define the overall (across all categories) virality score for image $h$ as

\vspace{-7pt}
\begin{equation}\label{Viral_rank}
V_{h}=\max_n A_h^n log\left(\frac{m_h}{\bar{m}}\right)
\end{equation}

where $m_h$ is the number of times image $h$ was resubmitted, and $\bar{m}$ is the average number of times any image has been resubmitted. 
If an image is resubmitted often, its virality score will be high. This ensures that images that became popular when they were posted, but 
were not reposted, are not considered to be viral (Fig.~\ref{fig:viral_pop_sample}). These often involve images where the content itself is 
less relevant, but current events draw attention to the image such as a recent tragedy, 
a news flash, or a personal success story e.g. ``Omg, 
I lost 40 pounds in 2 weeks''. On the other hand, images with multiple submissions seem more ``flexible'' 
for different titles about multiple situations and are arguably, intrinsically viral. Examples are shown in Fig.~\ref{fig:viral_exem}. 



\subsection{Viral Images Dataset}
\label{sec:10k_dataset}
\vspace{-5pt}


We use images from Reddit data collected in~\cite{Leskovec} to create our dataset. Lakkaraju \etal~\cite{Leskovec} crawled 
132k entries from Reddit over a period of 4 years. The entries often correspond to multiple submissions of the same image. 
We only include in our dataset images from categories (subreddits) that had at least 100 submissions 
so we have an accurate measure for $\bar{m}$ in Equation~\ref{Viral_rank}. 
We discarded animated GIFs. This left us with a total of 
10078 images from 20 categories, with $\bar{m}=6.7$ submissions per image.

We decided to use images from Reddit instead of other social networking sites such as Facebook 
and Google+~\cite{1309.3908} because users post images on Reddit  \textit{``\textsc{4thelulz}''} (i.e. just for fun) rather than personal 
social popularity \cite{berger2012makes}. 
We also prefer using Reddit instead of Flickr~\cite{www14_khosla} because 
images in Reddit are posted anonymously, hence they breed 
the purest form of ``internet trolling''. 


\subsection{Viral and Non-Viral Images Dataset}\label{viral_nonviral}
\vspace{-5pt}

Next, we create a dataset of 500 images containing the 250 most and least viral images each using Equation~\ref{Viral_rank}. 
This stark contrast in the virality score of 
the two sets of images gives us a clean dichotomy to explore as a first step in studying this complex phenomenon.
Recall that non-viral images include both -- images that did not get enough upvotes, and those that may have had many 
upvotes on one submission, but were not reposted multiple times. 

\subsubsection{Random Pairs Dataset}\label{viral_nonviral_random}
\vspace{-5pt}

In contrast with the clean dichotomy represented in the dataset above, we also create a dataset of pairs of images 
where the difference in the virality of the two images in a pair is less stark. We pair a random image from the 250 most viral images 
with a random image from $> 10k$ images with virality lower than the median virality. Similarly, we pair a random image from the 250 least viral 
images with a random image with higher than median virality. We collect 500 such pairs. Removing pairs that happen to have both images from top/bottom 
250 viral images leaves us with 489 pairs. We report our final human and computer results on this dataset, and refer to it as $(500_p)$ 
in Table~\ref{table:Method_Performance}. Training 
was done on the other 4550 pairs that can be formed from the remaining 10k images by pairing above-median viral images with below-median viral images.

\subsection{Viral Categories Dataset}\label{subreddit}
\vspace{-5pt}

For our last dataset, we work with the five most viral categories: funny, WTF, aww, atheism and gaming. 
We identify images that are viral only in one of the categories and not others. To do so, we compute the ratio between an 
image's virality scores with respect to the category that gave it the highest score among all categories that it was submitted to, and 
category that gave it the second highest score. That is, 

%

\vspace{-10pt}
\begin{equation}
\label{eq:ratio}
V^{c}_{h}=\frac{V^{{c}^1}_{h}}{V^{{c}^2}_{h}}
\end{equation}

where $V^{{c}^k}_{h}$ is the virality score image $h$ received on the category $c$ that gave it the $k^{th}$ highest score among all categories.

\vspace{-10pt}
\begin{equation}
V^{{c}^k}_{h}=A_h^{c^k}\pi\left(log\left(\frac{m_h^{c^k}}{\bar{m}_h}\right)\right)
\end{equation}

where $A_h^{n^k}$ is as defined in Equation~\ref{Jure_eq} for the categories that gave it the $k^{th}$ highest score among all 
categories that image $h$ was submitted to, $\pi(x)$ is the percentile rank of $x$, $m_h^{n^k}$ is the number of times image $h$ was submitted to 
that category, and $\bar{m}_h$ is the average number of times image $h$ was 
submitted to all categories. 
We take the percentile rank instead of the actual $log$ value 
to avoid negative values in the ratio in Equation~\ref{eq:ratio}.


To form our dataset, we only considered the top 5000 ranked viral images in our Viral Images dataset (Section~\ref{sec:10k_dataset}). 
These contained 1809 funny, 522 WTF, 234 aww, 123 atheism and 95 gaming images. Of these, we selected 85 images per category that had the highest 
score in Equation~\ref{eq:ratio} to form our Viral Categories Dataset.

\vspace{-5pt}
\section{Understanding Image Virality}\label{understanding}
\vspace{-5pt}

\begin{figure}[t]
\centering
%
\subfigure[WTF]{
    \includegraphics[scale=0.21,clip=true,draft=false,]{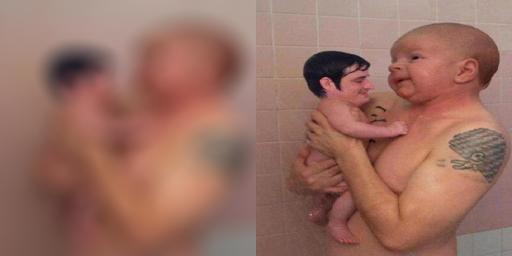}
    \label{fig:WTF_blur}
}
\subfigure[atheism]{
    \includegraphics[scale=0.21,clip=true,draft=false,]{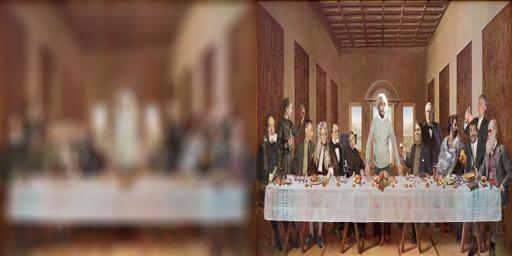}
    \label{fig:atheism_blur}
}
\vspace{-10pt}
\caption[]{Examples of temporal contextual priming through blurring in viral images. 
Looking at the images on the left in both \subref{fig:WTF_blur} and \subref{fig:atheism_blur}, what do you think the actual images depict? Did your expectations of the images turn out to be accurate?\vspace{-15pt}}
\label{fig:blur_answers}

\end{figure}

Consider the viral images of Fig.~\ref{fig:blur_answers}, where face swapping~\cite{bitouk2008face}, 
contextual priming~\cite{torralba2003contextual}, and scene gist~\cite{oliva2001modeling}
make the images quite different from what we might expect at a first glance. An analogous scenario researched in NLP is understanding the semantics of \textit{``That's what she said!''} jokes~\cite{kiddon2011s}.
We hypothesize that perhaps images that do not present such a visual challenge or contradiction -- 
where semantic perception of an image does not change significantly on closer examination of the image -- 
are ``boring''~\cite{leskovec2007dynamics,berger2012makes} and less likely to be viral. This contradiction 
need not stem from the objects or attributes within the image, but may also rise from the context of the image: be it the images 
surrounding an image, or the images viewed before the image, or the title of the image, and so on. Perhaps an interplay between these 
different contexts and resultant inconsistent interpretations of the image is necessary to simulate a visual double entendre leading to image virality.
With this in mind, we define four forms of context that we will study to explore image virality.

\begin{packed_enumerate}
 \item \textbf{Intrinsic context}: This refers to visual content that is intrinsic to the pixels of the image.
 \item \textbf{Vicinity context}: This refers to the visual content of images surrounding the image (spatial vicinity).
 \item \textbf{Temporal context}: This refers to the visual content of images seen before the image (temporal vicinity).
 \item \textbf{Textual context}: This non-visual context refers to the title or caption of the image. 
 These titles can sometimes manifest themselves as visual content (e.g. if it is photoshopped). A word graffiti 
 has both textual and intrinsic context, and will require NLP and Computer Vision for understanding.
\end{packed_enumerate}

\subsection{Intrinsic context}
\vspace{-5pt}

We first examine whether humans and machines can predict just by looking at an image, whether it is a viral image or not, and what the dominant 
topic (most suitable category) for the image is. 
For machine experiments, we use state-of-the-art image features such as DECAF6 deep features~\cite{donahue2013decaf}, gist~\cite{oliva2001modeling}, HOG~\cite{dalal2005histograms}, tiny images~\cite{torralba200880}, etc. using the implementation of~\cite{xiao2010sun}. 
We conduct our human studies on Amazon Mechanical Turk (AMT). We suspected that workers familiar with Reddit may have different 
performance at recognizing virality and categories than those unfamiliar with Reddit. So we created a qualification test that every worker had to take before 
doing any of our tasks. The test included questions about 
widely spread Reddit memes and jargon
so that anyone familiar with Reddit can easily get a high score, but workers who are not would get a very poor score. 
We thresholded this score to identify a worker as familiar with Reddit or not. 
Every task was done by 20 workers. Images were shown at 360 $\times$ 360. 

Machine accuracies were computed on the same test set as human studies.
Human accuracies are computed using a majority vote across workers. As a result (1) accuracies reported for different subsets of 
workers (e.g. those familiar with Reddit and those not) can each be lower than the overall accuracy, and (2) we can not 
report error bars on our results. We found that accuracies across workers on our tasks varied by $\pm 2.6\%$. On average, 73\% 
of the worker responses matched the majority vote response per image.

\vspace{-5pt}
\subsubsection{Predicting Topics}\label{subreddit_disc}
\vspace{-5pt}

We start with our topic classification experiment, where a practical application is to help a 
user determine which category to submit his image to.
We use our Viral Categories Dataset (Section~\ref{subreddit}). 
See Fig.~\ref{fig:montage_family_App} in Appendix. The images do generally seem distinct from one category to another. 
For instance, images that belong to the aww category seem to contain cute baby animals in the center of the image, images in atheism seem to have 
text or religious symbols, images in WTF are often explicit and tend to provoke feelings of disgust, fear and surprise.

\begin{figure}[t]
\centering
\subfigure[Category classification]{
    \includegraphics[scale=0.27,clip=true,draft=false,]{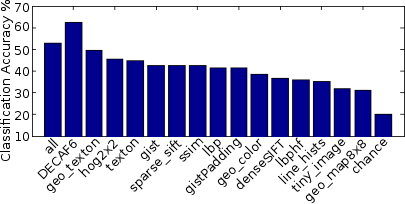}
    \label{fig:prediction_plot1}
}
\subfigure[Virality prediction]{
    \includegraphics[scale=0.27,clip=true,draft=false,]{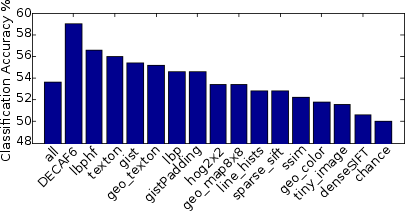}
    \label{fig:prediction_plot2}
}
\vspace{-10pt}
\caption[]{Machine accuracies on our Viral Categories (Section~\ref{subreddit}) and Viral \& Non-Viral Images 
datasets (Section~\ref{viral_nonviral}--tested on Top/Bottom 250 pairs), using different image features.\vspace{-15pt}
}

\label{fig:prediction}
\end{figure}

After training the 20 qualified workers with a sample montage of 55 images per category, 
they achieved  
a category identification accuracy of $87.84\%$ on 25 test images, 
where most of the confusion was between funny and gaming images. 
Prior familiarity with Reddit did not influence the accuracies because of the training phase. 
The machine performance using a variety of features can be seen in Fig.~\ref{fig:prediction_plot1}. A performance of 
$62.4\%$
 was obtained by using DECAF6~\cite{CloudCV}  (chance accuracy would be 20\%).  
 Machine and human confusion matrices can be found in Appendix II. 

\vspace{-10pt}
\subsubsection{Predicting Virality}\label{virality_prediction}
\vspace{-5pt}

Now, we consider the more challenging task of predicting whether an image is viral or not by looking at its content, by using 
our Viral and Non-Viral Images Dataset (Section~\ref{viral_nonviral}). 
We asked subjects on AMT whether they think a given image would be viral (i.e. ``become very viral on social networking websites like Facebook, Twitter, Reddit, Imgur, etc. with a lot of people liking, re-tweeting, sharing or upvoting the image?''). 
 Classification accuracy was $65.40\%$, where chance is $50\%$. 

In each of these tasks, we also asked workers if they had seen the image before, to get a sense for their bias based on familiarity with the image. 
We found that 
9\%, 1.5\% and 3\% of the images had been seen before by the Reddit workers, non-Reddit workers and all workers. While a small sample set, 
classification accuracies for this subset were high: $75.27\%$, $93.53\%$ and $91.15\%$. 
Note that viral images are likely to be seen even by non-Reddit users through other social networks.
Moreover, we found that workers who were familiar with Reddit in general had about the same accuracy as  workers who were not 
($63.24\%$ and $63.08\%$ respectively). They did however have different classification strategies. Reddit workers 
had a hit rate of $40.64\%$, while non-Reddit workers had a hit rate of
$28.96\%$. This means that Reddit workers were more likely 
to recognize an image as viral when they saw one (but may misclassify other non-viral images as viral). 
Non-Reddit workers were more conservative in calling images viral. 
Both hit rates under $50\%$ indicate a general bias towards labeling images as non-viral.
This may be because of the unnaturally uniform prior over viral and 
non-viral images in the dataset used for this experiment. 
Overall, workers who have never seen the image before and are not familiar with Reddit, 
can predict virality of an image better than chance. 
This shows that intrinsic image content is indicative of virality, and that image virality on communities
like Reddit is not just a consequence of snowballing effects 
instigated by chance.

%

\begin{figure}[t]
\centering
    \includegraphics[scale=0.23,clip=true,draft=false,]{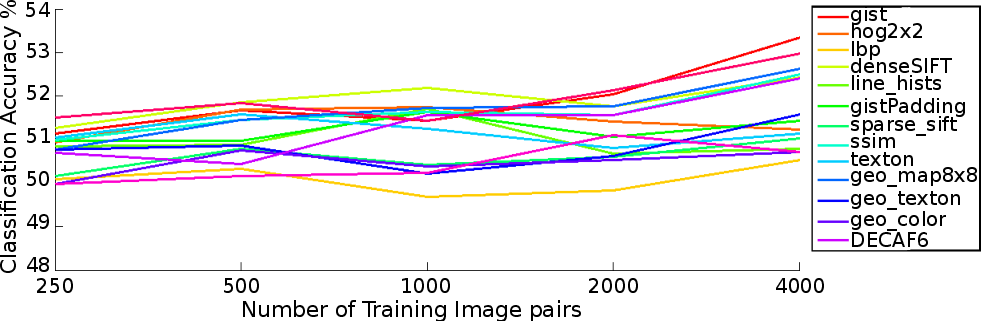}
    \label{fig:vm_metric}
\vspace{-7pt}
\caption[]
{Machine accuracy using our virality metric averaged across 5 random train/test splits, test set contained 2078 
random images each time. Notice that all descriptors produce chance like
results ($50\%$). Novel image understanding techniques need to be developed to predict virality.
\vspace{-20pt}
}
\label{fig:machine_score_prediction}

\end{figure}

Machine performance using our metric for virality is shown in Fig.~\ref{fig:machine_score_prediction}. Other metrics can be found in Appendix I. 
We see that current vision models have a hard time differentiating between these viral and non-viral images, under any criteria. 
The SVM was trained with both linear and non linear kernels 
on 5 random splits of our dataset of $\sim$10k images, using 250, 500, 1000, 2000, 4000 images for training, and 1039 images of each class for testing. 

The performance of the machine on the same set of images as used in the human studies using a variety of features to 
predict virality is shown in Fig.~\ref{fig:prediction_plot2}. 
Training was performed on the top and bottom 2000 images, excluding the top and bottom 250 images used for testing.
DECAF features achieve highest accuracy at 59\%;
This is above chance, but lower 
than human performance (65.4\%). The wide variability of images on Reddit (seen throughout the paper) and 
the poor performance of state-of-the-art image features indicates that automatic prediction of image virality will 
require advanced image understanding techniques. 

\vspace{-10pt}
\subsubsection{Predicting Relative Virality}
\vspace{-5pt}
\label{sec:relative_virality}

Predicting the virality of indivual images is a challenging task for both humans and machines. We therefore consider making relative predictions of virality.
That is, given a pair of images, 
is it easier to predict which of the two images is more likely to be viral? 
In psychophysics, this setup is called a two-alternative forced
choice (2AFC) task.

We created image pairs consisting of a random viral image and a random non-viral image from our Viral and Non-Viral Images 
dataset (Section~\ref{viral_nonviral}). We asked workers which of the two images is more likely to go viral. Accuracies were
all workers\footnote{62.12\% of AMT Workers were Reddit workers.}: $71.76\%$, Reddit workers: $71.68\%$ and 
non-Reddit workers: $68.68\%$, noticeably higher than $65.40\%$ on 
the absolute task, and 50\% chance. A SVM using DECAF6 image features got an accuracy of 
61.60\%, similar to the SVM classification accuracy on the absolute task (Fig.~\ref{fig:prediction_plot2}). 

\vspace{-10pt}
\subsubsection{Relative Attributes and Virality}
\vspace{-5pt}
\label{viral_attributes}

Now that we've established that a non-trivial portion of virality does depend on the image content, we wish to understand 
what kinds of images tend to be viral i.e. what properties of images are correlated with virality. We had subjects on AMT annotate the 
same pairs of images used in the experiment above, 
with relative attribute annotations~\cite{parikh2011relative}. 
In other words, for each pair of images, we asked them which image has more of an attribute presence 
than the other. 
Each image pair thus has a 
relative attribute annotation $\in\{-1,0,+1\}$ indicating whether the first image has a stronger, equal or weaker presence of the 
attribute than the second image. 
In addition, each image pair has a $\in\{-1,+1\}$ virality annotation based on our 
ground truth virality score indicating whether the first image is more viral or the second. 
We can thus compute the correlation between each relative attribute and relative virality.

\begin{figure}[t]
\centering
\subfigure[Correlations of human-annotated attributes with virality]{
\includegraphics[scale=0.285,clip=true,draft=false,]{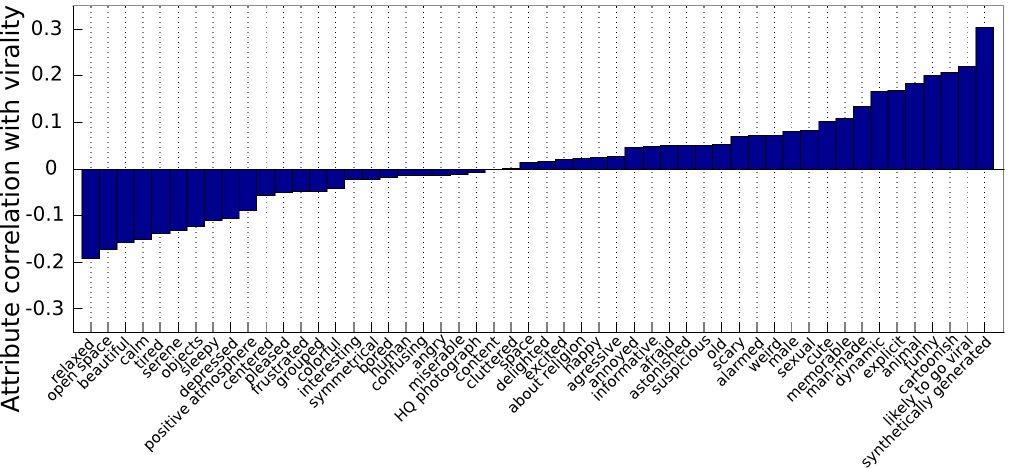}
    \label{fig:relative_graph}}
\subfigure[Correlation of attribute combinations with virality ($>5000$ pairs). The Force condition puts tiebreakers on neutral atts.]{
\includegraphics[scale=0.19,clip=true,draft=false,]{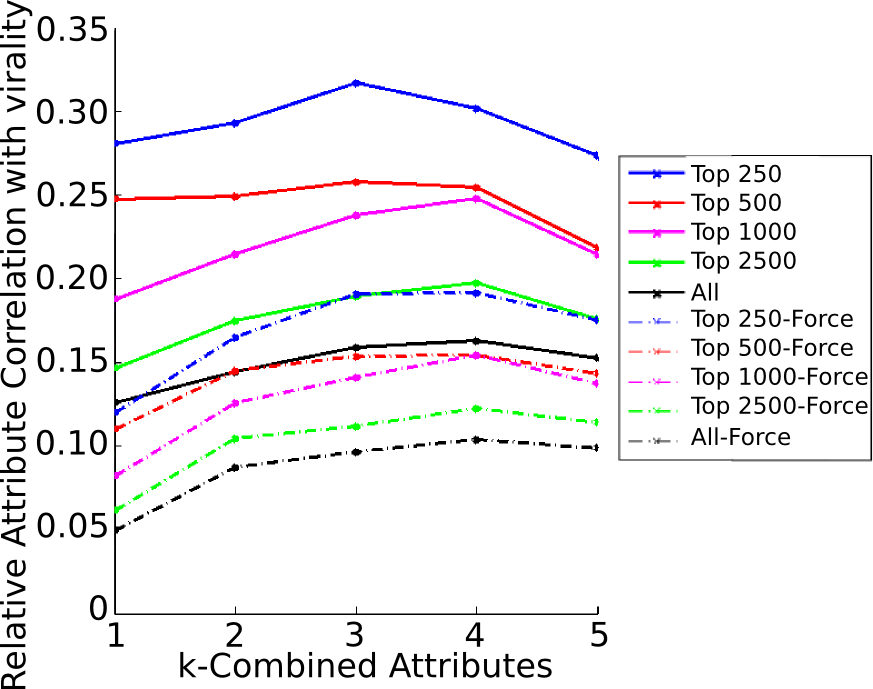}
    \label{fig:Optimal_Behaviour}}
    \hspace{10pt}
\subfigure[Correlation of attribute combinations with virality after priming (Top/Bottom $250$ pairs: Section~\ref{viral_nonviral})]{
\includegraphics[scale = 0.26]{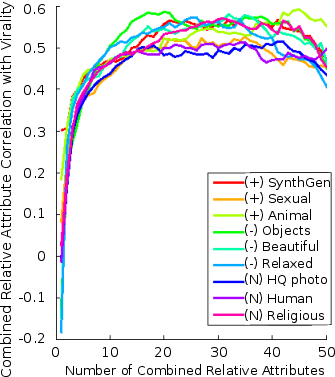}
\label{fig:relative_performance}}
\vspace{-20pt}
\caption{The role of attributes in image virality.\vspace{-20pt}}
\end{figure}

We selected 52 attributes that capture the spatial layout of the scene, the aesthetics of the image, the subject of the image,
how it made viewers feel, whether it was photoshopped, explicit, funny, etc. Inspirations for these attributes came from familiarity with Reddit, 
work on understanding image memorability~\cite{isola2011understanding}, and representative emotions on the valence/arousal 
circumplex~\cite{berger2011arousal,guerini2015deep}. See Fig.~\ref{fig:relative_graph} for the entire list of attributes we used. 
As seen in 
Fig.~\ref{fig:relative_graph}, synthetically generated (Photoshopped), cartoonish and funny images are most likely to be viral, while 
beautiful images that make people feel calm, relaxed and sleepy (low arousal emotions~\cite{berger2011arousal}) are least likely to be viral. 
Overall, correlation values between any individual attribute and virality is low, due to the wide variation in the kinds of 
images found on communities like Reddit.

We further studied virality prediction with combinations of attributes.
We start by identifying the single (relative) attribute with the highest (positive or negative) correlation 
with (relative) virality. We then greedily find the second attribute that when added to the first one, increases 
virality prediction the most.
For instance, funny images tend to be viral, and images with animals tend to be viral. 
But images that are funny \emph{and} have animals may be even more likely to be viral. 
The attribute to be added can be the attribute itself ($\uparrow$), 
or its negation ($\downarrow$). This helps deal with attributes that are negatively correlated with virality. For instance, synthetically generated images that 
are \emph{not} beautiful are more likely to be viral than images that are either synthetically generated or not beautiful. In this way, we greedily 
add attributes. Table~\ref{table:Combined_Relative_Attributes} shows the attributes that 
collaborate to correlate well with virality. We exclude ``likely to go viral'' and ``memorable'' from 
this analysis because those are high-level concepts in themselves, and would not add to our understanding of virality.

A combination of 38 attributes 
leads to a virality predictor that achieves an
accuracy of $81.29\%$. This can be viewed as a hybrid human-machine predictor of virality. The attributes 
have been annotated by humans, but the attributes have been selected via statistical analysis. We see that this significantly 
outperforms humans alone ($71.76\%$) and the machine alone ($59.00\%$, see Table~\ref{table:Method_Performance}). 
One could train a 
classifier on top of the attribute predictors to further boost performance, but the semantic interpretability provided by 
Table~\ref{table:Combined_Relative_Attributes} would be lost. Our analysis begins to give us an indication of which image properties 
need to be reliably predicted to automatically predict virality.

We also explore the effects of ``attribute priming'': if the first attribute in the combination is one 
that is negatively correlated with virality, how easy is it to recover from that to \emph{make} the image viral? 
Consider the scenario where an image is very ``relaxed'' (inversely correlated with virality). Is it possible for a graphics designer to induce 
virality by altering other attributes of the image to make it viral? Fig.~\ref{fig:relative_performance} shows the 
correlation trajectories as more attributes are greedily added to a ``seed'' attribute that is positively $(+)$, 
negatively $(-)$, or neutrally $(N)$ correlated with virality. We see that in all these scenarios, an image can be made viral 
by adding just a few attributes. Table~\ref{table:Combined_Relative_Attributes} lists which attributes are selected for 3 
different ``seed'' attributes.
Interestingly, while sexual is positively 
correlated with virality, when seeded with animal, not sexual increases 
the correlation with virality. As a result, when we select our five attributes greedily, 
the combination that correlates best with virality is:  animals, synthetically generated, not beautiful, 
explicit’’ and \emph{not} sexual.

\vspace{-10pt}
\subsubsection{Automated Relative Virality Prediction}
\vspace{-5pt}

To create an automated relative virality prediction classifier, we start by using our complete $\sim$10k image dataset and have AMT workers
do the same task as in Section~\ref{viral_attributes}, by dividing them into viral (top half in rank) vs non viral (lower half in rank), and
randomly pairing them up for relative attribute annotation 
for the top 5\footnote{Tagging all 52 relative attributes accurately for all $5k$ image pairs in the dataset is expensive.}
performing attributes from our greedy search in Fig.~\ref{fig:relative_performance}: 
Animal, Synthetically Generated(SynthGen), Beautiful, Explicit and Sexual. Note that all of our top-5 attributes are visual. 
Correlation trajectories of combined attributes for all our dataset in a hybrid human-machine virality predictor can be seen at Fig.~\ref{fig:Optimal_Behaviour}.  

\begin{table}[t]
\tiny
\centering 
\scriptsize{
\resizebox{\textwidth}{!}{%
\begin{tabular}[H]{|c||c|c|c|c|c|}
\cline{1-3}
\Xhline{2\arrayrulewidth} 
 & 1 & 2 & 3 & 4 & 5 \\ \cline{1-6}
\Xhline{2\arrayrulewidth}
Attribute (+) & $\uparrow$ synth. gen. & $\uparrow$ animal & $\downarrow$ beautiful & $\uparrow$ explicit & $\downarrow$ sexual \\ \cline{1-6}
Virality Correlation & 0.3036 & 0.3067 & 0.3813 & 0.3998 & 0.4236 \\ \cline{1-6}
\Xhline{3\arrayrulewidth}
Attribute (-) & $\uparrow$ beautiful & $\uparrow$ synth. gen. & $\uparrow$ animal & $\uparrow$ dynamic & $\uparrow$ annoyed \\ \cline{1-6}
Virality Correlation & -0.1510 & 0.2383 & 0.3747 & 0.3963 & 0.4097 \\ \cline{1-6}
\Xhline{3\arrayrulewidth}
Attribute (N) & $\uparrow$ religious & $\uparrow$ synth. gen. & $\uparrow$ animal & $\downarrow$ beautiful & $\uparrow$ dynamic \\ \cline{1-6}
Virality Correlation & 0.0231 & 0.1875 & 0.3012 & 0.3644 & 0.3913 \\ \cline{1-6}
\Xhline{2\arrayrulewidth}
\end{tabular}}
\vspace{-5pt}
{\caption{Correlation of human-annotated attribute combinations with virality. Combinations are ``primed'' with the first attribute.\vspace{-10pt}}\label{table:Combined_Relative_Attributes}}
}
\end{table}


\begin{table}[t]
\centering
\renewcommand\footnoterule{}
\scriptsize{
\begin{tabular}[H]{|c|c|c|}
\cline{1-3}
Dataset & Classification Method & Performance \\ \cline{1-3}
\Xhline{2\arrayrulewidth}
 & Chance & $50\%$ \\ \cline{1-3} 
\Xhline{2\arrayrulewidth}
All images & SVM + image features & $53.40\%$ \\ \cline{2-3}
\Xhline{2\arrayrulewidth}
 & Human (500) & $71.76\%$ \\ \cline{2-3}
Top/Bottom & SVM + image features $(500)$ & $61.60\%$ \\ \cline{2-3}
250 viral & Human annotated Atts.-1 $(500)$ & $56.77\%$ \\ \cline{2-3}
(Section~\ref{viral_nonviral}) & Human annotated Atts.-3 $(500)$ & $68.53\%$ \\ \cline{2-3}
 & Human annotated Atts.-5 $(500)$ & $71.47\%$ \\ \cline{2-3}
 & Human annotated Atts.-11 $(500)$ & $73.56\%$ \\ \cline{2-3}
 & Human annotated Atts.-38 $(500)$ & $\mathbf{81.29\%}$ \\ \cline{2-3}
\Xhline{2\arrayrulewidth}
Top/Bottom & Khosla~\etal Popularity API~\cite{www14_khosla} $(500_p)$ & $51.12\%$ \\ \cline{2-3}
250 viral & SVM + image features $(500_p)$ & $58.49\%$ \\ \cline{2-3}
paired with & Human $(500_p)$ & $60.12\%$\\ \cline{2-3}
random imgs. & Human annotated Atts.-5 $(500_p)$& $65.18\%$ \\ \cline{2-3}
(Section~\ref{viral_nonviral_random}) & SVM + Deep Attributes-5 $(500_p)$ & $\mathbf{68.10\%}$ \\ \cline{1-3} 
\end{tabular}
}
\vspace{-10pt}
\captionof{table}{Relative virality prediction across different datasets \& methods. \vspace{-15pt}}
\label{table:Method_Performance}
\end{table}

With all the annotations, we then train relative attribute predictors for each of these attributes with DECAF6 deep features~\cite{donahue2013decaf} 
and an SVM classifier   
through 10-fold cross validation to obtain relative attribute predictions on all image pairs (Section ~\ref{viral_nonviral_random}). The relative attribute 
prediction accuracies we obtain are:
Animal: $70.14\%$, Synthgen: $45.15\%$, Beautiful: $56.26\%$, Explicit: $47.15\%$, Sexual: $49.18\%$ (Chance: $33.33\%$),
by including neutral pairs. Futhermore, we get 
Animal: $87.91\%$, Synthgen: $67.69\%$, Beautiful: $81.73\%$, Explicit: $65.23\%$, Sexual: $71.13\%$ for $+/-$ relative labels, excluding neutral (tied) 
pairs (Chance: $50\%$). Combining these automatic attribute predictions to inturn (automatically) predicted virality, 
we get an accuracy of $68.10\%$. 
If we use ground truth relative attribute annotations for these 5 attributes 
we achieve ($65.18\%$) accuracy, better than human performance ($60.12\%$) at predicting relative virality directly from images.
Using our deep relative attributes, machines can predict relative virality more accurately than humans! 
This is because (1) humans do not fully understand what makes an image viral (hence the need for a study like this and 
automatic approaches to predicting virality) and (2) the attribute classifiers trained by the machine may have latched on to biases 
of viral content. The resultant learned notion of attributes may be different from human perception of these attributes. 


Although our predictor works well above chance, notice that extracting attributes from these images is non-trivial, given the diversity of 
images in the dataset. While detecting faces and animals is typically considered to work reliably enough~\cite{girshick2013rich}, recall that 
images in Reddit are challenging due to their non-photorealism, embedded textual content and image composition. 
To quantify the qualitative difference in the images in typical vision datasets and our dataset, we trained a classifier to 
classify an image as belonging to our Virality Dataset or the SUN dataset~\cite{xiao2010sun,torralba2011unbiased}. We extracted DECAF6 features 
from our dataset and similar number of images from the SUN dataset. The resultant classifier was able to classify a new image as coming 
from one of the two datasets with $90.38\%$ accuracy, confirming qualitative differences. 
Moreover, the metric developed for popularity~\cite{www14_khosla} applied to our dataset outputs chance like results (Table~\ref{table:Method_Performance}).
Thus, our datasets provide a new regime to study image understanding problems.


\subsection{Vicinity context}
\vspace{-5pt}


\begin{figure*}[t]
\centering
\begin{tabular}{ccccc}
\subfigure[car pair\label{fig:img_pair_1}]{
\includegraphics[scale=0.11]{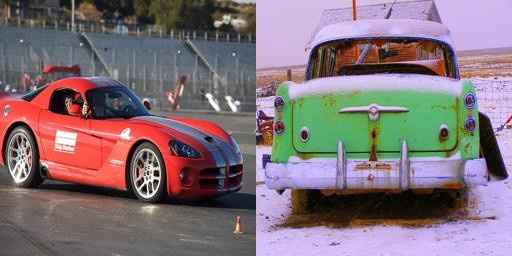}
    } &
\multirow{-2}[2.5]{*}{\subfigure[car set]{\includegraphics[scale=0.09]{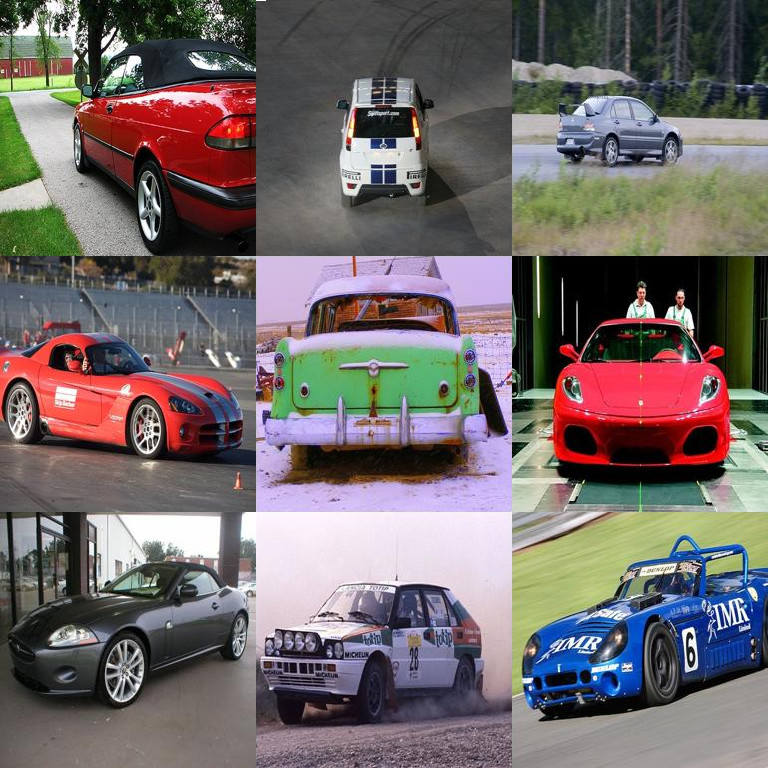}    \label{fig:img_complete_1}}} &
\multirow{-2}[2.5]{*}{\subfigure[set saliency]{\includegraphics[scale=0.09]{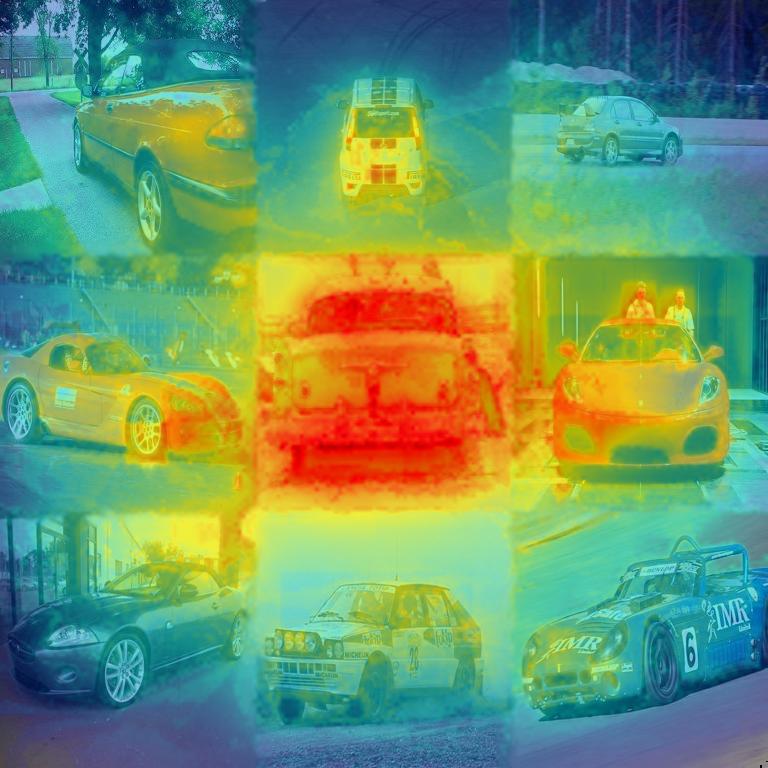}\label{fig:img_complete_2}}} &
\multirow{-2}[2.5]{*}{\subfigure[Sort 4]{\includegraphics[scale=0.28]{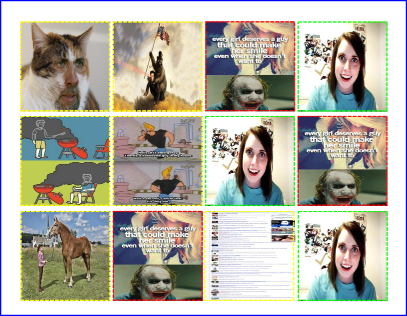}\label{fig:mash1}}} &
\multirow{-2}[2.5]{*}{\subfigure[Sort 2]{\includegraphics[scale=0.28]{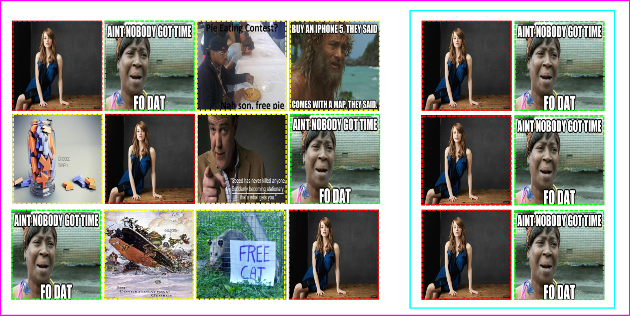}\label{fig:mash2}}} \\
\subfigure[pair saliency]{\includegraphics[scale=0.11]{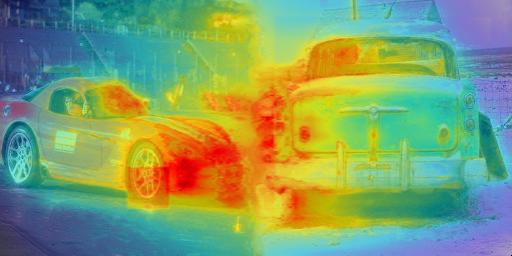}
    \label{fig:img_pair_2}}\\
\end{tabular}
\vspace{-10pt}
\caption[]{The value of how red a car is, 
or whether one car is more red than the other~\subref{fig:img_pair_1} does not change if more
images are added to the pool~\subref{fig:img_complete_1}. However, an image that may seem more viral - visualized through saliency~\cite{judd2009learning} 
(e.g. the red vintage Ferrari in~\subref{fig:img_pair_2}) than another image, may start seeming less viral than the same image depending on the images 
added to the mix. See Fig.~\ref{fig:affinity}~\subref{fig:img_complete_2}. In our experiments, workers are asked to sort four images in ascending order of their virality in one experimental 
design~\subref{fig:mash1}, while they are asked to sort only 2 images in another design~\subref{fig:mash2}, after being shown all 4 of them. 
In both cases, there are only two target 
images of interest (viral:green, non-viral:red), while the other two images are proxy images (yellow) added to the mix.  
These images are chosen such that they are close (in gist space) to the viral target image (top row), the non-viral target image (middle row), 
or random (bottom row).
\vspace{-15pt}
}
\label{fig:affinity}
\end{figure*}

Reasoning about pairs of images as we did with relative virality above, leads to the question of the impact of images in the vicinity of an image 
on human perception of its virality. We designed an AMT experiment to explore this (Fig.~\ref{fig:affinity}). Recall that in the previous experiment involving relative virality prediction, we formed pairs of images, 
where each pair contained a viral and non-viral image. We now append these pairs with two ``proxy'' images. These proxies are selected to be either similar to the 
viral image, or to the non-viral image, or randomly. Similarity is measured using the gist descriptor~\cite{oliva2001modeling}. The $4^{th}$ and $6^{th}$ most 
similar images are selected from 
our Viral Images dataset (Section~\ref{sec:10k_dataset}). We do not select the two closest images to avoid near 
identical matches and to ensure that the task did not seem like a ``find-the-odd-one-out'' task. 
We study these three conditions in two different experimental settings. 
The first is where workers are asked to sort all four images from what they believe is the least viral to the most viral.
In the second experimental design, workers were still shown all four images, but were asked to only annotate which one of the two images from the original 
pair is more viral than the other. Maybe the mere presence of the ``proxy'' images affects perception of virality?
For both cases, we only check the relative ranking of the viral and non viral image.

\begin{wraptable}{r}{4.7cm}
\vspace{-10pt}
\scriptsize{
\begin{tabular}{|c|c|c|c|}   \toprule
& {Sort 4} & {Sort 2} \\ \midrule
{Viral-NN } & $65.16\%$ & $66.64\%$  \\  \midrule
{Non viral-NN} & $68.60\%$  & $65.56\%$  \\  \midrule
{Random} & $\mathbf{52.24\%}$ & $65.00\%$ \\  \bottomrule
\end{tabular}
}
\vspace{-10pt}
\captionof{table}{Human ranking accuracy across different proxy images.}
\label{table:KNN_Classification}
\end{wraptable}

Worker accuracy in each of the six scenarios is shown in Table~\ref{table:KNN_Classification}. 
We see that when asked to sort all four images, identifying the true viral images is harder 
with the presence of random proxies, as they tend to confuse workers and their performance at predicting virality drops to nearly chance. The presence of carefully selected proxies can still make the target viral image salient.
When asked to sort just the two images of interest, 
performance is overall higher (because the task is less cumbersome). But more importantly, performance is very similar across the 
three conditions (Sort 2). This suggests that perhaps the mere presence of the proxy images does not impact virality prediction. 

Developing group-level image features that can reason about such higher-order phenomenon has not been well studied in the 
vision community. Visual search or saliency has been studied to identify which images or image regions pop out. But models of change in relative 
orderings of the same set of images based on presence of other images have not been explored. Such models may allow us to select the ideal set of images 
to surround an image by to increase its chances of going viral. 

\subsection{Temporal context}\label{temporal_context}
\vspace{-5pt}

Having examined the effect of images in the spatial vicinity on image virality, we now study the effects of temporal aspects. In particular, 
we show users the same pairs of images used in the relative virality experiment in Section~\ref{sec:relative_virality} at 4 different 
resolutions one after the other: $8 \times 8$, $16 \times 16$, $32 \times 32$, $360 \times 360$ (original). We choose blurring to simulate first impression judgements
at thumbnail sizes when images are `previewed'. At each stage, we asked them which 
image they think is more likely to be viral. Virality prediction performance was 
$47.08\%$, $49.08\%$, $51.28\%$ and $62.04\%$.
Virality prediction is reduced to chance even in $32 \times 32$ images, where humans have been shown to recognize semantic content in images 
very reliably~\cite{torralba200880}. Subjects reported being surprised for 65\% of the images. We found a -0.04 correlation between true virality and surprise,
and a -0.07 correlation between predicted virality and surpise. Perhaps people are bad at 
estimating whether they were truly surprised or not, and asking them may not be effective; or surprise truly is not correlated with virality.

\subsection{Textual context}
\vspace{-5pt}

As a first experiment to evaluate the role of the title of the image, we show workers pairs of images and ask them which one
they think is more likely to be viral. We then reveal the title of the image, and ask them the same question again. 
We found that access to the title barely improved virality prediction 
($62.04\%$ vs. $62.82\%$).
This suggests that perhaps the title does not sway subjects after they have already judged the content.

Our second experiment had the reverse set up. We first showed workers the title alone, and asked them which title is 
more likely to make an image be viral. We then showed them the image (along with the title), and asked them the same question. Workers' 
prediction of relative virality was worse than chance using the title alone ($46.68\%$). Interestingly, having been primed by the title, even with access to the image performance did not improve significantly 
above chance ($52.92\%$) and is significantly lower than their performance when viewing an image without being primed by the title ($62.04\%$). 
This suggests that image content seems to be the prime signal in human perception of image virality. 
However, note that these experiments do not analyze the role of text that may be embedded in the image (memes!).

\vspace{-5pt}
\section{Conclusions}
\vspace{-5pt}

We studied viral images from a computer vision perspective. We introduced three new
image datasets from Reddit, the main engine of viral content around the world. We defined a virality score 
using Reddit metadata. We found that virality can be predicted more accurately as a relative concept. While humans can predict 
relative virality from image content, machines are unable to do so using low-level features. High-level image understanding is key. We identified 
five key visual attributes that correlate with virality: Animal, Synthetically Generated, (Not) Beautiful, Explicit and Sexual. We predict these 
relative attributes using deep image features. Using these deep relative attribute predictions as features, machines (SVM) can predict virality with an 
accuracy of $68.10\%$ 
(higher than human performance: $60.12\%$). Finally, we study how human prediction of image virality varies with different ``contexts'' -- 
intrinsic, spatial (vicinity), temporal and textual. 
This work is a first step in understanding the complex but important 
phenomenon of image virality. We have demonstrated the need for advanced image understanding to predict virality, as well as the qualitative difference 
between our datasets and typical vision datasets. This opens up new opportunities for the vision community. 
Our datasets and annotations will be made publicly available.


\vspace{-5pt}
\section{Acknowledgements}
This work was supported in part by ARO YIP 65359NSYIP to D.P. and NSF IIS-1115719. We would also like to thank Stanislaw Antol, Michael Cogswell, Harsh Agrawal, 
and Arjun Chandrasekaran for their feedback and support.
\vspace{-5pt}

\vspace{-5pt}
\section*{Appendix I: Virality Metrics}\label{App1}

\begin{figure}[!h]
\centering
\subfigure[Virality metric: $V_{h}$]
{
    \includegraphics[scale=0.23,clip=true,draft=false,]{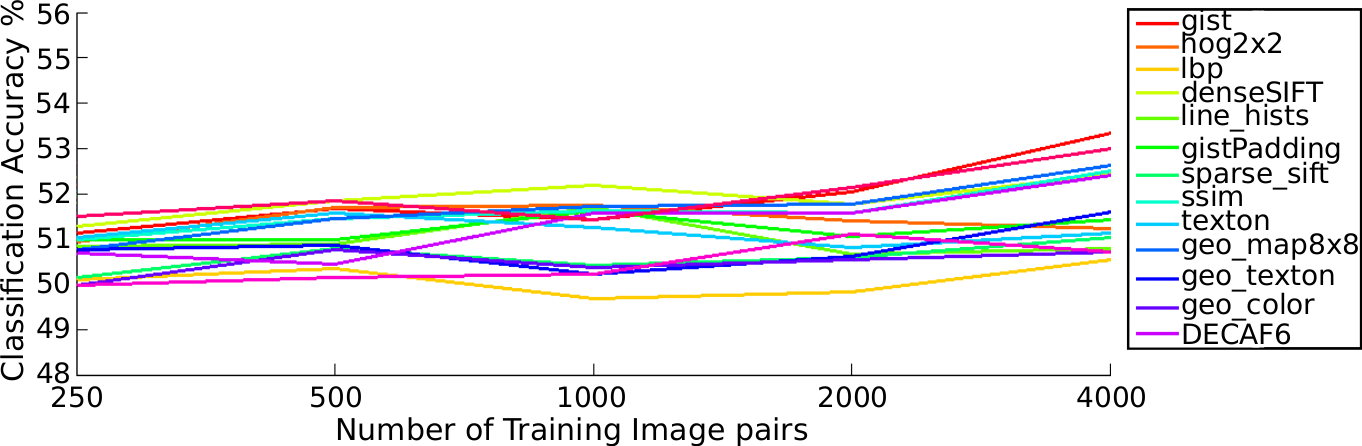}
    \label{fig:vm_metric_App}
}

\subfigure[Maximum upvotes: $\max_n\{A_h^n\}$]{
    \includegraphics[scale=0.23,clip=true,draft=false,]{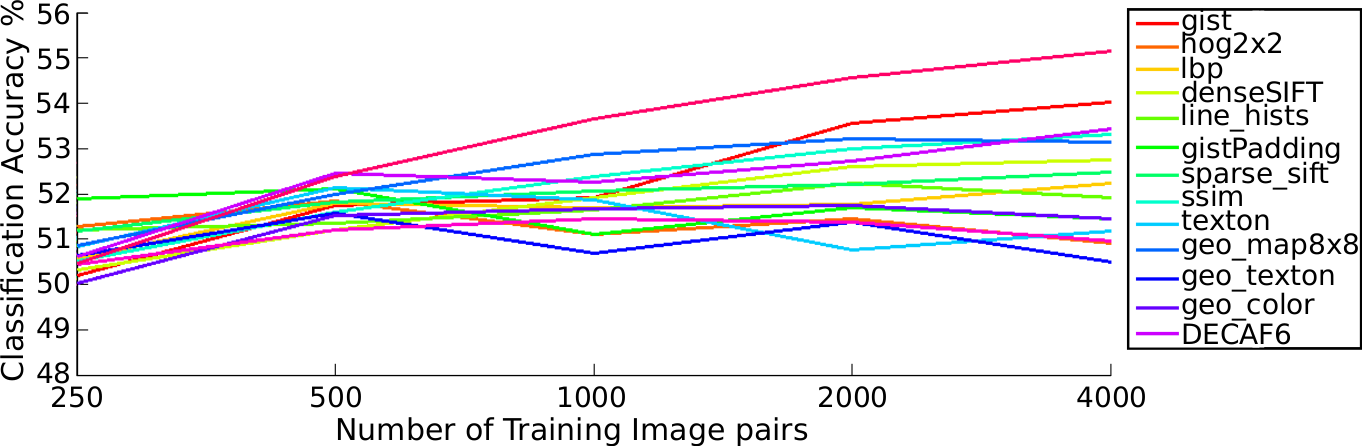}
    \label{fig:up_metric_App}
}

\subfigure[Number of resubmissions: $m_h$]{
    \includegraphics[scale=0.23,clip=true,draft=false,]{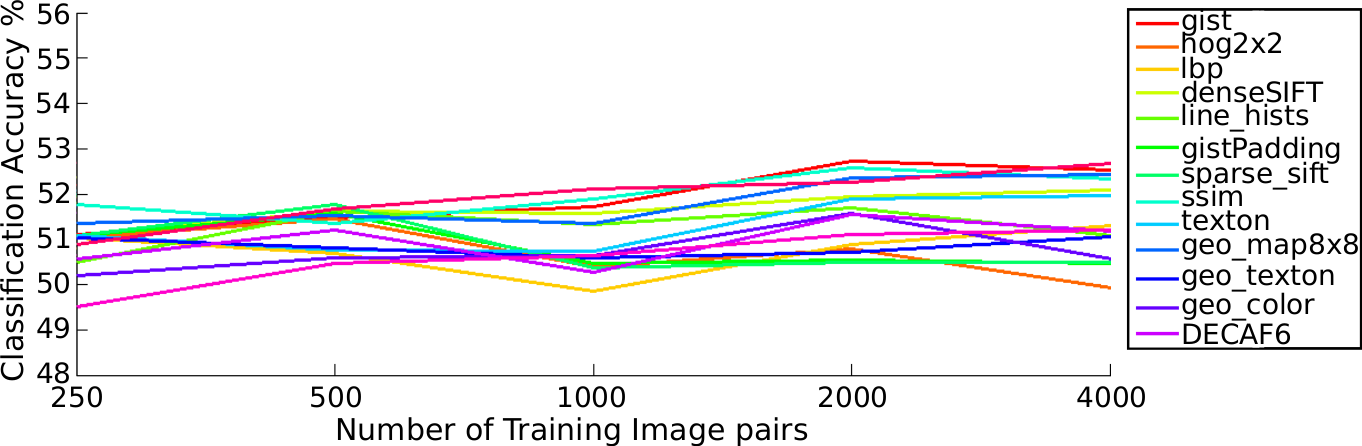}
    \label{fig:sh_metric_App}
}
\vspace{-7pt}
\caption[]
{Machine accuracy using our virality metric~\ref{fig:vm_metric_App}, and other metrics~\ref{fig:up_metric_App},~\ref{fig:sh_metric_App}. Notice 
that all descriptors produce chance like
results. Novel image understanding techniques need to be developed to predict virality.
}
\label{fig:machine_score_prediction_App}
\end{figure}

Machine performance using different metrics for virality are shown in Fig.~\ref{fig:machine_score_prediction_App}. 
We see that current vision models have a hard time differentiating between these viral and non-viral images, under any criteria. 
The SVM was trained with both linear and non linear kernels 
on 5 random splits of our dataset of $\sim$10k images, using 250, 500, 1000, 2000, 4000 images for training, 
and 1039 images of each class for testing.

Recall that we define our virality score for image $h$ across all resubmissions $n$ as

\vspace{-7pt}
\begin{equation}\label{Viral_rank}
V_{h}=\max_n A_h^n log\left(\frac{m_h}{\bar{m}}\right)
\end{equation}

where $m_h$ is the number of times image $h$ was resubmitted, and $\bar{m}$ is the average number of times any image has been resubmitted. $A_h^n$ is a normalized
score based on the metric of~\cite{Leskovec}. The other metrics
we experimented with included maximum upvotes $\max_n \{A_h^n\}$(Fig.~\ref{fig:up_metric_App}) across submissions and 
number of resubmissions $m_h$(Fig.~\ref{fig:sh_metric_App}).

\begin{figure}[!t]
\centering
\subfigure[AMT workers]{
    \includegraphics[scale=0.20,clip=true,draft=false,]{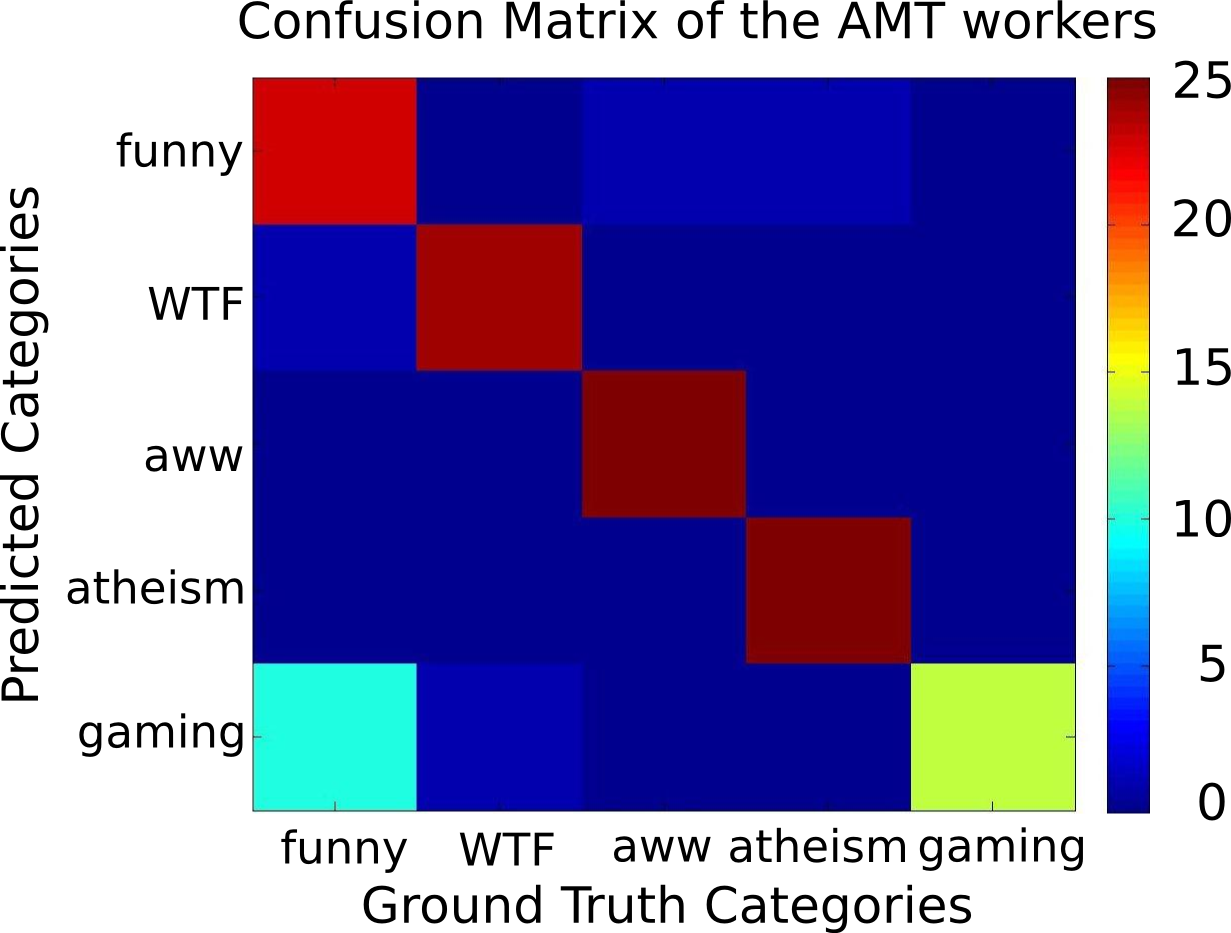}
    \label{fig:AMT_confusion_App}
}
\subfigure[SVM-DECAF6]{
    \includegraphics[scale=0.20,clip=true,draft=false,]{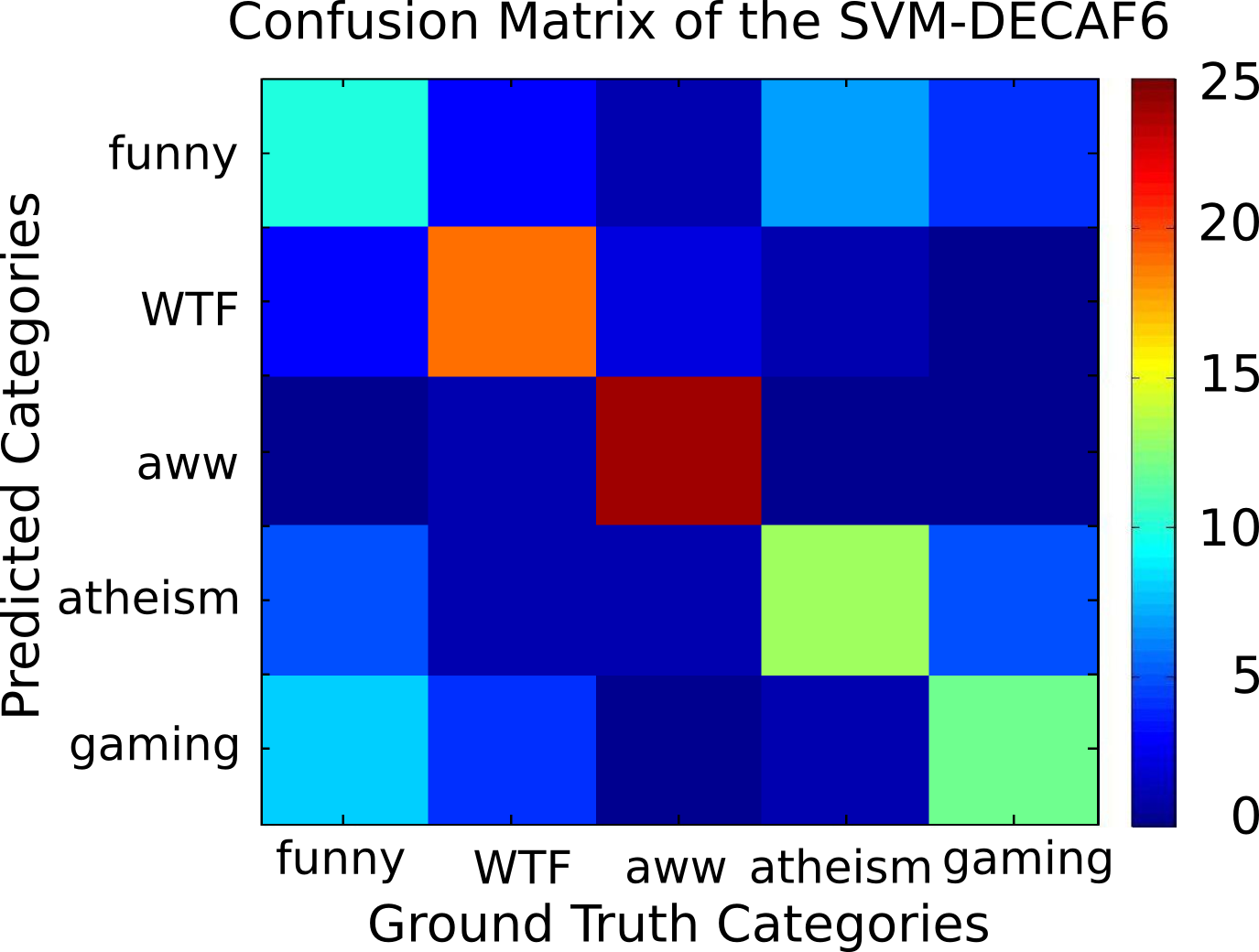}
    \label{fig:SVM_confusion_App}
}
\vspace{-10pt}
\caption[]{Confusion matrices in \subref{fig:AMT_confusion_App}, \subref{fig:SVM_confusion_App} for topic (subreddit) classification show that humans \& machines can do a reasonable job of 
recognizing the dominant topic of a viral image. Both find aww to be the easiest to recognize 
and both tend to confuse gaming images with funny. 
There are $90.4\%$ of the images correctly classified by either the machine or humans ($9.6\%$ are misclassified by both).
}
\label{fig:subreddit_exem_App}
\end{figure}

\section*{Appendix II: Category Prediction Confusion Matrices}

The confusion matrices in Figure~\ref{fig:subreddit_exem_App} depict how a human and computer perform at identifying categories/topics (subreddits) in viral images (Section~\ref{subreddit_disc}). A 
computer achieves $62.4\%$ accuracy using SVM + DECAF6 after training on 60 images,
while a human achieves $87.84\%$ on the same test images (chance is $20\%$). Recall that this group of AMT workers had to undergo training to learn how to identify the categories
just like the computer, with the same exemplars (55 on training  + 5 for validating that the workers paid attention during training). Both humans and machines tend to confuse gaming images with funny ones. They
are also both remarkabely good at identifying the aww category. While for a human it might be trivial, we suspect that it is also easy for the machine, 
since texture (animal fur, feathers, skin) plays an important role that DECAF6 is encoding. Atheism on the other hand is very simple for a human, but complex for 
a machine. We are inclined to believe that since 
identifying whether an image contains religious content or not involves high-level semantics and often text, 
the task of detecting atheism is challenging for a machine. The machine as a result tends to confuse it with
funny or gaming, as some of these images also contain text. 
Examples from all 5 categories are shown in Fig~\ref{fig:montage_family_App}.

\begin{figure*}[!t]
\centering
\subfigure[SUN dataset]
{
    \includegraphics[scale=0.5,clip=true,draft=false,]{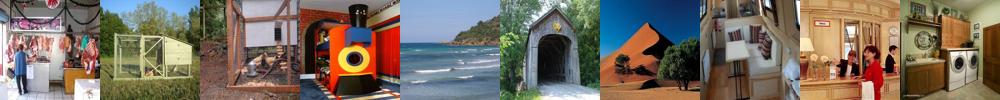}
    \label{fig:SUN_mini_App}
}
\subfigure[Flickr dataset used in~\cite{www14_khosla} for image popularity]{
    \includegraphics[scale=0.5,clip=true,draft=false,]{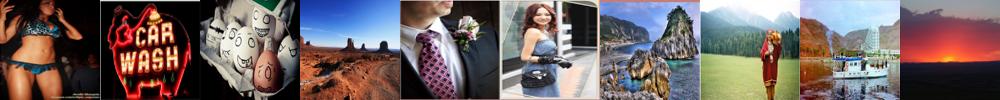}
    \label{fig:Flickr_mini_App}
}

\subfigure[Reddit dataset used in our work for image virality. Virality scores are shown in the bottom of each image. `Philosoraptor' a celebrity meme, 
scores remarkably higher than other synthetic images.] 
{
    \includegraphics[scale=0.625,clip=true,draft=false,]{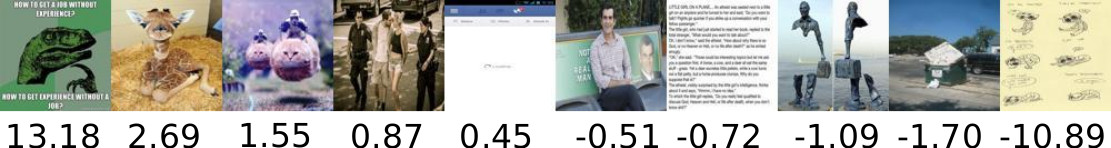}
    \label{fig:Reddit_mini_App}
}
\vspace{-7pt}
\caption[]
{There is a qualititative difference in each dataset~\cite{torralba2011unbiased}. 
Notice that the images in~\ref{fig:SUN_mini_App} and~\ref{fig:Flickr_mini_App} are still constrained to photographs of everyday objects and scenes.
However, a majority of images in~\ref{fig:Reddit_mini_App} are highly complex: they include text, cartoons, and 
out-of-context objects. In our dataset, this is the norm, rather than the exception. 
}
\label{fig:dataset_montage_App}
\end{figure*}

Potential applications for subreddit classification include a recomendation system for a human to know to what category to submit his image for marketing/personal
purposes. 
Automatically identifying a subreddit (or topic) of the image can also provide context when generating image descriptions~\cite{Ordonez:2011:im2text,karpathy2014deep}.

%

\section*{Appendix III: Dataset Comparison}

Viral images are very different from standard images analyzed in different computer vision datasets. In Figure~\ref{fig:SUN_mini_App}, we see how the SUN dataset images
favors open spaces in outdoor and indoor environments. Images from Flickr (Figure~\ref{fig:Flickr_mini_App}) have more variation than SUN images, yet the images still follow many traditional 
and photographic rules: rule of the third, principal subject(s) in picture, all co-occurring objects are in context, and they are colorful just like the SUN dataset. Recall that Flickr 
has become an online platform for photographers and amateurs to share their most beautiful pictures, an attribute that correlates \textit{inversely} with virality.

Viral images, on the other hand, seem non-sensical, chaotic, and very unpredictable. They do not follow photographic rules. They can have cartoons, text,
and (non)photorealistic content embedded separately or together, yet they still express an idea that is easily understandable for humans. 
We hope that the vision community will study automatic image understanding in these kinds of images (Figure~\ref{fig:Reddit_mini_App}).

\vspace{-15pt}
\begin{figure}[!t]
\centering
\includegraphics[scale=0.24,clip=true,draft=false,]{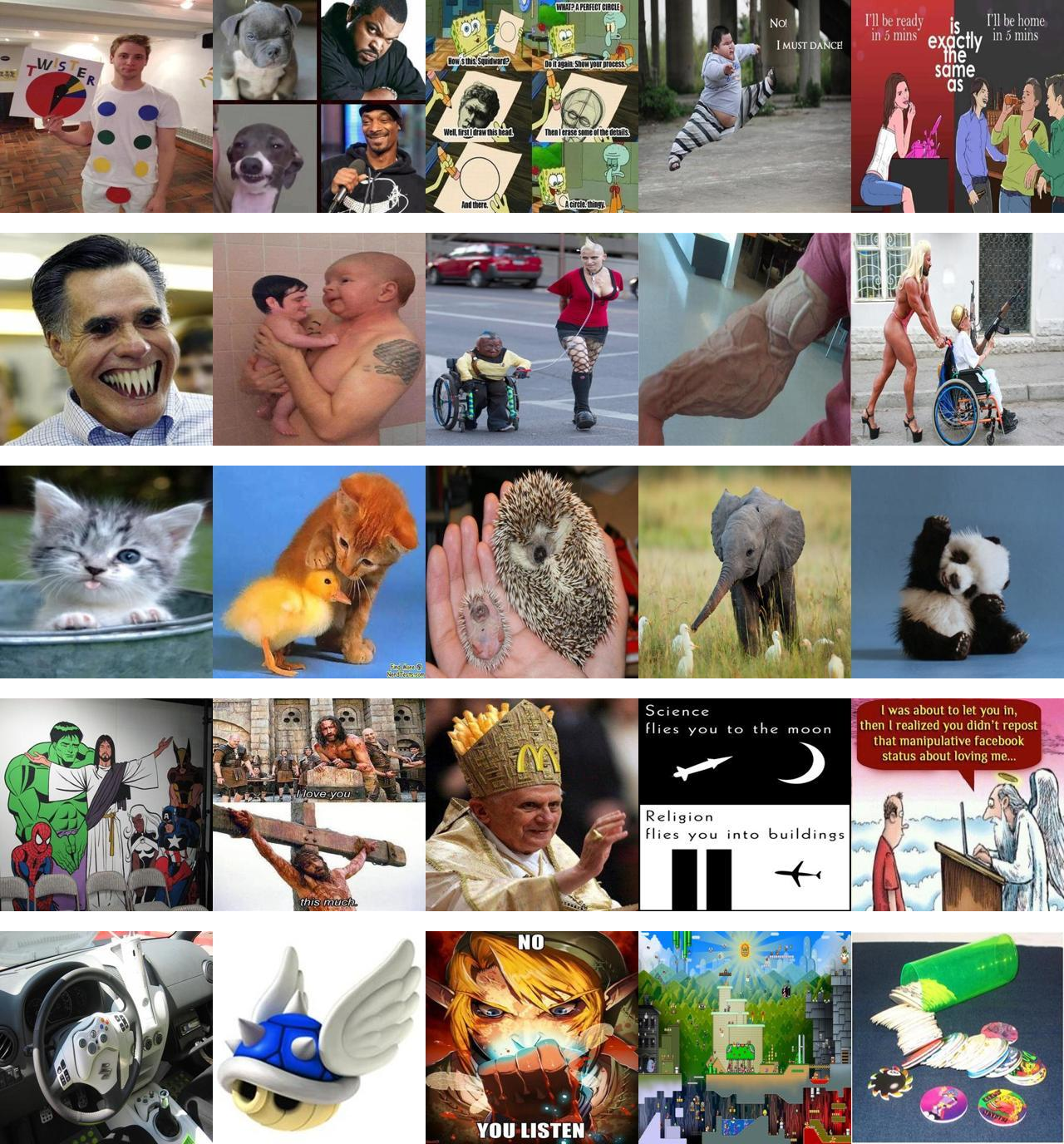}
\label{fig:montage_family_App}
\vspace{-10pt}
\caption[]{
Example images from the 5 most viral categories (top to bottom): funny, WTF, aww, atheism, gaming.
}
\label{fig:montage_family_App}
\end{figure}


\clearpage
\clearpage
{\footnotesize
\bibliographystyle{ieee}
\bibliography{egbib_fix}
}
\end{document}